\documentclass[twoside]{article}
\usepackage{fleqn,espcrc2}

\usepackage[dvips]{graphicx, color}

\newcommand{\ncm}{\newcommand}
\newcommand{\rencm}{\renewcommand}
 
\def\f{\varphi}    

\def\l{\lambda}
\def\m{\mu}

\def\o{\omega}
\def\p{\pi}       

\def\D{\Delta}

\ncm{\dsp}{\displaystyle}
\ncm{\nn}{\nonumber}
\ncm{\nnn}{\nonumber\linebreak[4]}
\ncm{\nit}{\noindent}
\ncm{\del}{\partial}
\ncm{\av}[1]{\mbox{$\langle #1 \rangle$}}
\ncm{\ket}[1]{\mbox{$| #1 \rangle$}}
\ncm{\bra}[1]{\mbox{$\langle #1 |$}}
\ncm{\avc}[1]{\mbox{$\langle #1 \rangle_{\psi}$}}
\ncm{\half}{\mbox{{\small $\frac{1}{2}$}} }
\ncm{\quart}{\mbox{{\small $\frac{1}{4}$}} }
\ncm{\tq}{\mbox{{\small $\frac{3}{4}$}} }
\ncm{\third}{\mbox{{\small $\frac{1}{3}$}} }
\ncm{\sixth}{\mbox{{\small $\frac{1}{6}$}} }
\ncm{\eigth}{\mbox{{\small $\frac{1}{8}$}} }
\ncm{\thrhalf}{\mbox{{\small $\frac{3}{2}$}} }
\ncm{\thrfor}{\mbox{{\small $\frac{3}{4}$}} }
\ncm{\twothi}{\mbox{{\small $\frac{2}{3}$}} }
\ncm{\fivtwo}{\mbox{{\small $\frac{5}{2}$}} }

\ncm{\dxx}{\mbox{$\partial_{x}^2$}}
\ncm{\dx}{\mbox{$\partial_{x}$}}
\ncm{\dt}{\mbox{$\partial_{t}$}}
\ncm{\dtt}{\mbox{$\partial_{t}^2$}}
\ncm{\un}{1\!\!1}
\ncm{\RE}{\mbox{Re}}
\ncm{\IM}{\mbox{Im}}
\ncm{\Tr}{\mbox{tr}\,}
\ncm{\diag}{\mbox{diag}\,}
\ncm{\Det}{\mbox{Det}\,}
\ncm{\Log}{\mbox{Log}\,}
\ncm{\ra}{\rightarrow}
\ncm{\la}{\leftarrow}
\ncm{\dg}{\dagger}
\ncm{\pr}{\prime}
\ncm{\ha}{\hat{a}}
\ncm{\hP}{\hat{P}}

\ncm{\aplt}{ \mbox{}_{\textstyle \sim}^{\textstyle < }     }
\ncm{\apgt}{ \mbox{}_{\textstyle \sim}^{\textstyle > }     }
\ncm{\Oa}{\mbox{$\mbox{O}(a)$}}
\ncm{\Sp}{\mbox{\hspace{1.0cm}}}
\ncm{\capit}[1]{\caption{\it #1}}

\def\be{\begin{equation}}
\def\ee{\end{equation}}
\def\bea{\begin{eqnarray}}
\def\eea{\end{eqnarray}}
\def\bi{\begin{itemize} \itemsep = 0.01\itemsep  }
\def\bii{\begin{itemize} \small \itemsep = 0.01\itemsep }
\def\ei{\end{itemize}}
\def\bc{\begin{center}}
\def\ec{\end{center}}
\def\bs{\begin{slide}}
\def\es{\end{slide}}

\def\beac{\begin{eqnarray} \color [rgb] {0,0,1} }
\def\eeac{\end{eqnarray} }

\ncm{\shead}[1]{\bc { \Large \color [rgb]{1.0, .0, .0} #1 \normalcolor} \ec}
\ncm{\ssubh}[1]{{\large \color [rgb]{1.0,.0,.1} #1 \normalcolor}}

\rencm{\thefootnote}{\mbox{\protect{$\fnsymbol{footnote}$}} }

\ncm{\front}[5]   
{
   \begin{titlepage}
      \noindent {#1} \hfill {#2}\\
      \begin{center}
         \vspace{1.5\baselineskip}
         {\Large\bf  #3  } \\
         \vspace{2\baselineskip}
         \vspace{1.5\baselineskip}
          #4\\
         \vspace{1.5\baselineskip}
   
         Institute for Theoretical Physics, University of Amsterdam, \\
         Valckenierstraat 65, 1018 XE Amsterdam,
         The~Netherlands.
    
      \end{center}
      \vfill
      {\bf Abstract}\\
       #5
   \end{titlepage} 
}

\ncm{\frontslide}[4]
{
   \begin{titlepage}
      \noindent {#1} \hfill {#2}\\
      \begin{center}
         \vspace{1.5\baselineskip}
         {\Large\bf  #3  } \\
         \vspace{2\baselineskip}
         \vspace{1.5\baselineskip}
          #4\\
         \vspace{1.5\baselineskip}
   
         Institute for Theoretical Physics, \\
         Valckenierstraat 65, 1018 XE Amsterdam,
         The~Netherlands.
    
      \end{center}
   \end{titlepage} 
}

\begin{document}

\title{ New Initial Conditions for Quantum Field Simulations after a Quench }

 \author{ M.~Sall\'e, J.~Smit and J.C.~Vink\thanks{ 
 Presented by J.~Vink 
 at the Lattice 2001 conference, 19-24 August, 2001, Berlin, Germany.
 } \\
Institute for Theoretical Physics, Valckenierstraat 65, 
1018 XE Amsterdam, The Netherlands
}


\begin{abstract}

We investigate a new way of using the quantum fluctuations in the vacuum as 
initial conditions for subsequent classical field dynamics. We show that with 
this method the field properly thermalizes at later times, whereas the method 
used previously leads to unphysical results.

\end{abstract}

\maketitle

\section{Initial conditions for a quench}

Investigation of initial value problems for quantum fields using numerical
simulation are difficult.
The dynamics cannot be computed exactly and one has to use e.g.\ the classical
approximation, the large $N$ or Hartree approximation. 
It is then not obvious how to specify initial conditions properly.
Here we focus on this problem and investigate a new way to supply
initial conditions for classical dynamics. We shall apply the same strategy
to a simulation with the recently proposed \cite{SaSm00} Hartree 
ensemble approximation as well.

As a test case, we consider a quench in a simple $1+1$ dimensional
$\f^4$ model discretized on a lattice, with hamiltonian,
\be
  H   =  \sum_x [ \half \p_x^2 - \half  \f_x\D\f_x  
             + \half \m^2\f_x^2 + \quart \l \f_x^4.
   \label{VPOT}
\ee
We model a sudden quench by assuming that at $t=0$ the sign of $\m^2$ in the
potential flips from positive to negative. At $t=0$
the system is in the vacuum state corresponding to the single-well
potential ($\m^2>0$). Classically this would imply that the field is at rest,
but triggered by (vacuum) quantum fluctuations, the field modes with momentum
$|p|<|\m|$ will grow exponentially in a ``spinodal decomposition'' of the field.
It is not obvious how one should supply quantum fluctuations as initial
conditions for the classical fields. Until now one has followed the prescription
\cite{iniAll} that the 
classical field at $t=0$ should reproduce the one- and two-point functions 
evaluated in the quantum theory. 

This implies that initial values for the (complex) Fourier 
amplitudes $\f_p$ and $\p_p$ of the classical field and its velocity 
$\p = \dot\f$, must be drawn from a gaussian ensemble with probability
distribution (using $\o_{0p} = (p^2 + |\m^2|)^{1/2}$),
\vspace{-0.7mm}
\be
  P(\f,\p) \propto \exp[ -\sum_p ( |\p_p|^2 + \o_{0p}^2 |\f_p|^2)/2\o_{0p} ],
  \label{PVAC}
\ee
\vspace{-0.7mm}

An apparent problem with this prescription is that the initial vacuum 
fluctuations contribute to the energy density and the equilibrium temperature
that would correspond to the amount of energy put into the field diverges
when the cut-off is removed, $T \propto 1/a$. One can only hope that in the 
early stages of the field evolution, this reservoir of spurious energy in the
short wavelength modes does not significantly affect the low-energy
dynamics. 

These problems are absent in the new approach we shall describe next.
During the spinodal decomposition, there may be a time window $t_- < t < t_+$ 
in which $\av{ \o_p |\f_p|^2 }\gg \half$ while 
 the backreaction can still be ignored, 
$\quart\l \f^4 \ll \half |\m^2| \f^2$.  If this is so, one can 
solve the quantum initial value problem analytically
in the approximation $\l = 0$ for $t \le t_0$ with $t_- < t_0 < t_+$.
At time $t_0$ we switch to a classical simulation since occupation
numbers (of the low momentum modes) have grown sufficiently large.
As matching conditions we impose that the one- and two-point functions
with {\em subtracted} vacuum fluctuations, are reproduced in the ensemble of 
classical initial conditions. 

Using the $\l=0$ solution, we can compute the one- and two-point functions,
$\av{\f_p}  = \av{\p_p} = 0$, $ \av{|\f_p|^2}$ and $\av{|\p_p|^2}$.
The two-point functions contain
contributions from vacuum fluctuations which we would like to subtract.
Even though the quench model at early times is clearly far out of equilibrium,
we nonetheless use the following equilibrium relations to {\em define} 
the (time dependent) particle energy $\o_p$ and number density $n_p$,
\be
 \av{ |\f_p|^2 } = (n_p + \half)/\o_p,\; \av{ |\p_p|^2 } = (n_p + \half)\o_p.
  \label{INICOR}
\ee
In this way we identify the contribution from vacuum fluctuations by the
$\half$ in these relations.
To switch to classical dynamics, we specify a gaussian ensemble 
of classical fields at $t=t_0$ with 
$ \av{|\f^{\rm cl}_p|^2} = n_p/\o_p$ and $\av{|\p^{\rm cl}_p|^2} = n_p\o_p$.
Even though the analytical calculation gives a tail $n_p \propto 1/p^4$ for
large momenta, shall make the further simplification that $n_p=0$ for the 
$|p|>\m$.

For comparison we shall also show results, still using classical dynamics, but 
using the customary \cite{iniAll} initial ensemble that includes vacuum 
fluctuations, $n_p \ra n_p+\half$.
In this case one might argue that the coupling constants need to be 
renormalized, and that somehow the effect of the vacuum fluctuations should
be subtracted. As this is a classical theory which is far from
equilibrium, it is not clear how to implement such a
renormalization in a consistent way.

Finally we use the Hartree ensemble approximation to compute the dynamics.
Now we must specify initial conditions for the mean fields and for
the mode functions. In order to be as close to classical dynamics as possible,
we choose an ensemble of initial conditions, in which the mean fields have
the same distribution as the classical fields above at time $t_0$.
Vacuum fluctuations are represented by the
mode functions, which we choose as plane waves.
We refer to ref.\ \cite{SaSm00} for further details of the Hartree 
ensemble method.  

\section{Thermalization}

\begin{figure} [tb]
\bc
\hspace{-0.7cm}
\scalebox{0.50}[0.66]{ \includegraphics[clip]{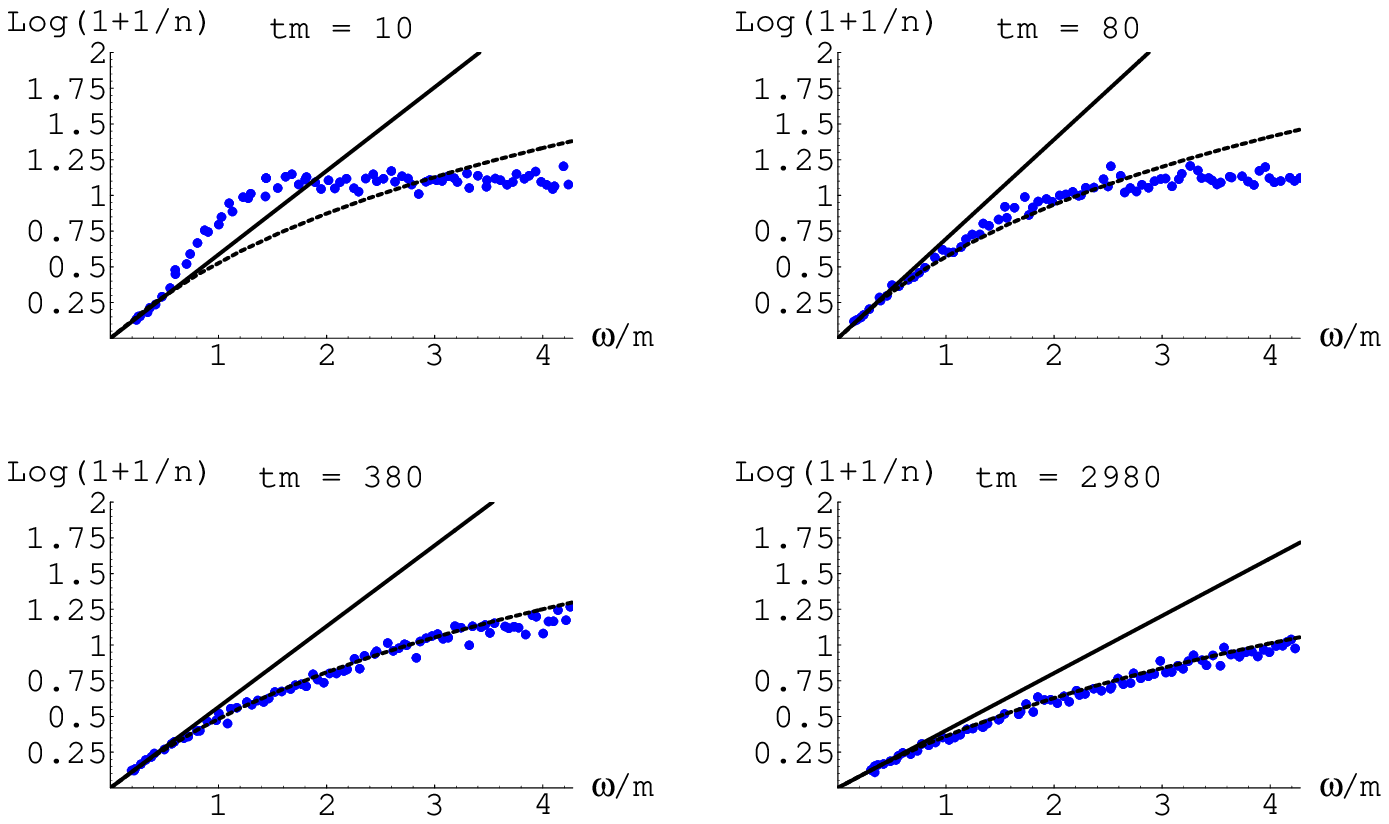} }
\vspace{-1.1cm}
\capit{
Particle number density using classical dynamics with
initial conditions that include vacuum fluctuations.
The model parameters are: $\l/m^2 = 1/12$, $Lm = 102.4$,
$1/am = 5$.
\label{FIGold}
}
\ec
\vspace{-1.1cm}
\end{figure}

\begin{figure} [tb]
\bc
\hspace{-0.7cm}
\scalebox{0.50}[0.66]{ \includegraphics[clip]{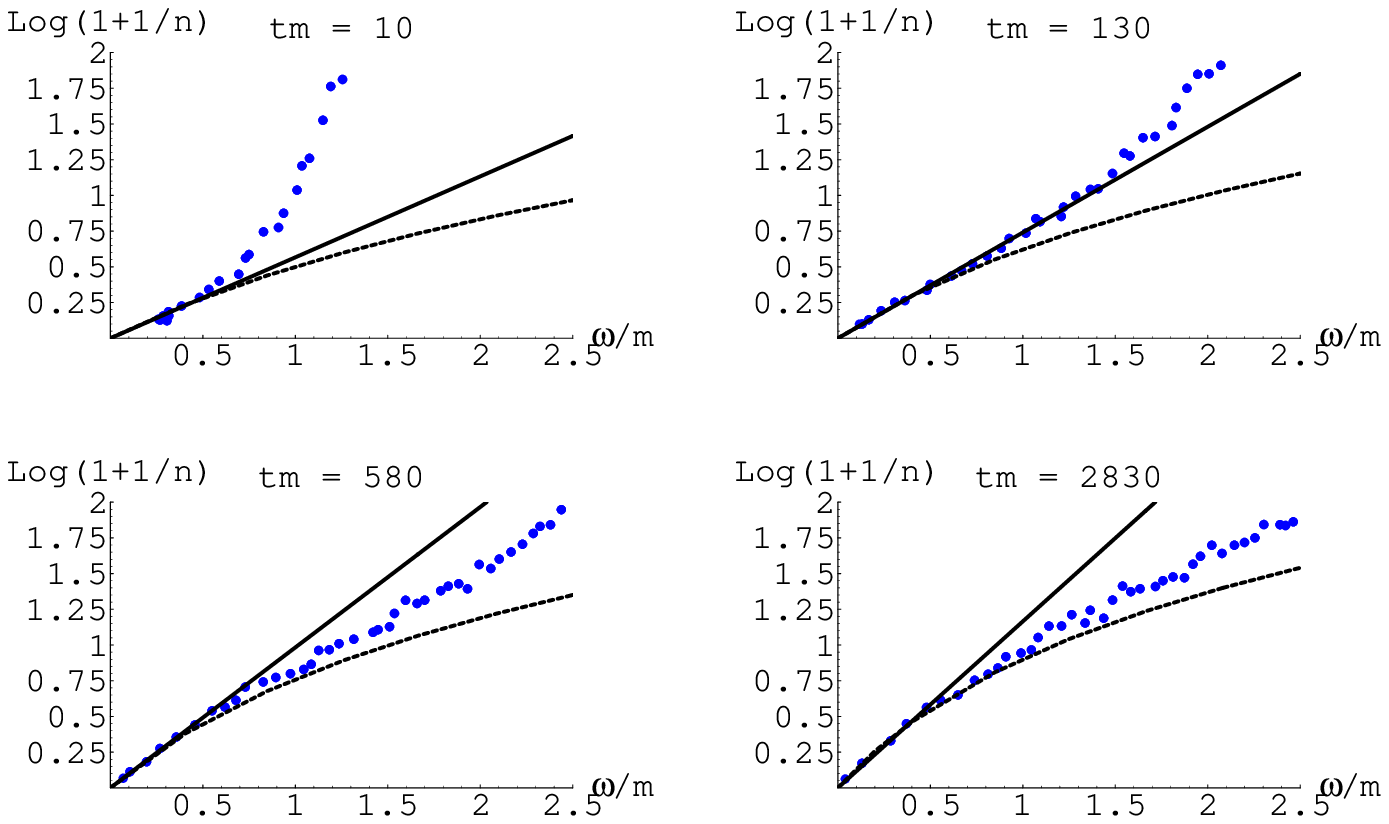} }
\vspace{-1.0cm}
\capit{
The same as Fig.\ \ref{FIGold} but using initial conditions {\em without} 
vacuum fluctuations.
\label{FIGnew}
}
\ec
\vspace{-1.0cm}
\end{figure}

\begin{figure} [tb]
\bc
\hspace{-0.7cm}
\scalebox{0.66}[0.62]{ \includegraphics[clip]{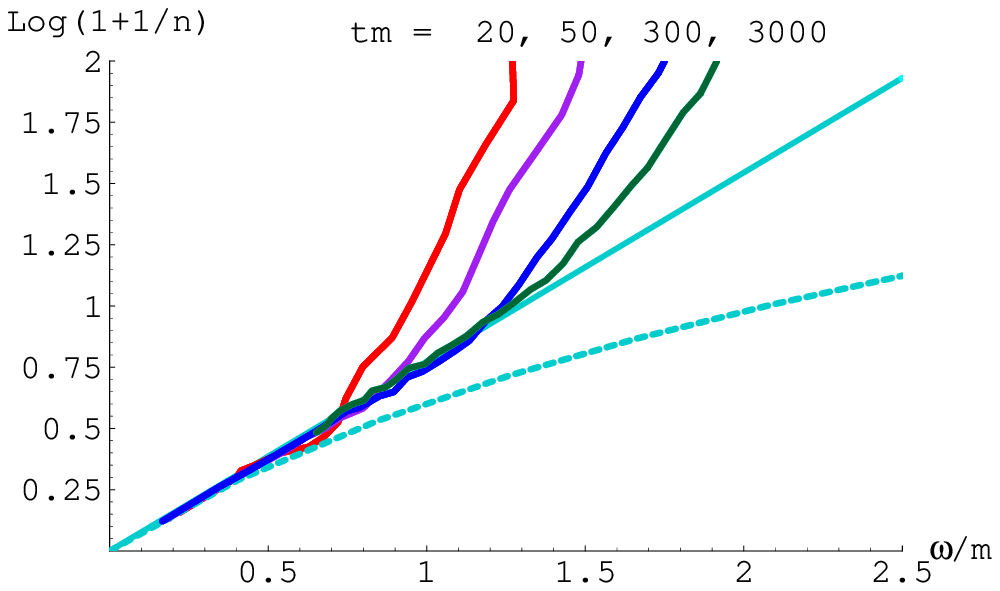} }
\vspace{-1.0cm}
\capit{
The same as Figs.\ \ref{FIGold}  and \ref{FIGnew} but using the Hartree
ensemble approximation.
\label{FIGhartree}
}
\ec
\vspace{-1.0cm}
\end{figure}

In order to study the effect of the different initial conditions, we
monitor the thermalization process. Hence we plot the time dependence
of the particle number density which, for classical fields, is
defined as in (\ref{INICOR}) but without subtracting the $\half$,
$ n_p = ( \av{|\f_p|^2}\av{|\p_p|^2})^{1/2} $.
 In Figs.\ \ref{FIGold}--\ref{FIGhartree}
we show $\log(1+1/n)$, which is linear in $\o$ for a Bose-Einstein distribution.
We switch from the
analytical solution to the numerical method at $ t_0m = 0.6$;
the results do not change when this time is chosen differently, provided
it stays in the range $0.2\aplt t_0m \aplt 2$.

In Fig.\ \ref{FIGold} we use classical dynamics with the customary initial 
conditions 
including high momentum vacuum fluctuations, as in ref.\ \cite{iniAll}
At very early time $tm=5-10$, one recognizes the unstable modes from the
large values of $n$ at $\o \aplt 0.5$. Also the vacuum fluctuations are
clearly visible, since they imply that $n\approx \half$ or 
$\log(1+1/n)\approx 1.1$.

\begin{figure} [tb]
\bc
\hspace{-0.7cm}
\scalebox{0.66}[0.62]{ \includegraphics[clip]{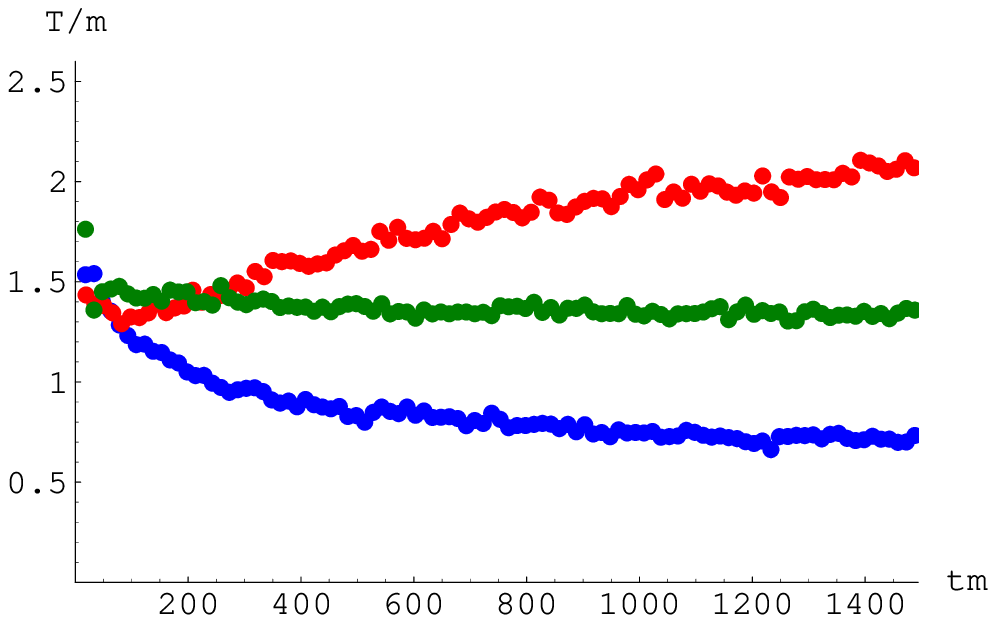} }
\vspace{-1.0cm}
\capit{
Time dependence of the temperatures obtained from the
particle densities at low $\o$ of Figs.\ \ref{FIGold} (top), 
\ref{FIGnew} (bottom) and \ref{FIGhartree} (middle).
\label{FIGtemps}
}
\ec
\vspace{-1.0cm}
\end{figure}

At later times modes with larger momenta thermalize with
a classical distribution, indicated by the curved line $n_p = T/\o_p$.
Eventually, at $tm\apgt 3000$ all modes have thermalized and the corresponding
temperature has become unphysically large. The temperature we obtain from
fitting the (low $\o$) part of the spectrum, 
is plotted in Fig.\ \ref{FIGtemps}. After an initial decrease one
clearly sees the unphysical rising of the temperature for $tm\apgt 75$.

Next we use the new initial conditions, which do not include vacuum 
fluctuations.  As can be seen in Fig.\ \ref{FIGnew}, the low-momentum
modes at early times behave similarly to the previous case.
But now the number density is exponentially suppressed for large momenta.
At lager times we see that
the distribution gradually moves towards a classical shape, but this
happens much more slowly than in Fig.\ \ref{FIGold}.


The lack of spurious vacuum energy in the simulation of Fig.\ \ref{FIGnew}
also has a pronounced effect on the temperature we obtain by fitting $n$
against $T/\o$ for low $\o$. 
Now the initial temperature drop continues, as energy continues to 
dissipate from the low momentum modes towards higher momenta.

Finally we use the Hartree ensemble method to compute the field dynamics.
For increasing time, more and more modes thermalize and align towards a
Bose-Einstein distribution. At the largest time that we simulated, $tm=3000$
the particle numbers follow a BE distribution over a significant range
of energies.

The temperature, obtained from fitting the low-momentum part of the
spectrum with a BE-distribution, is shown in Fig.\ \ref{FIGtemps}.
At very early times, $t\aplt 50$, the behavior appears to be the same as in 
the previous classical simulations.
At later times the temperature decreases very slowly, much
more slowly than with the classical dynamics. This is understandable because
the particle number densities at high momenta remain exponentially suppressed.

\section{Conclusion}

Using vacuum fluctuations as seeds for subsequent classical dynamics has
to be done with care. The straightforward way, choosing
an initial classical ensemble that reproduces the quantum two-point functions,
leads to misleading results at late times: the temperature of low-energy 
particles that equilibrate keeps rising and will eventually reach an unphysical
value $T\propto 1/a$. 
We have further demonstrated that the new prescription for initial conditions
gives quite reasonable results, both when used with classical and with
Hartree ensemble dynamics.

\end{document}